\documentclass{pasj00}
\draft

\begin{document}
\SetRunningHead{V. Bak{\i}\c{s} et al.}{Young Detached Triple System LT CMa}
%\Received{}%{yyyy/mm/dd}
%\Accepted{}%{yyyy/mm/dd}
%\Published{}%{yyyy/mm/dd}

\title{Absolute Dimensions and Apsidal Motion of the Young Detached System LT Canis Majoris}

\author{%
Volkan \textsc{Bak{\i}\c{s}}\altaffilmark{1}
\.{I}brahim \textsc{Bulut}\altaffilmark{1}
Sel\c{c}uk \textsc{Bilir}\altaffilmark{2}
Hicran \textsc{Bak{\i}\c{s}}\altaffilmark{1}
Osman \textsc{Demircan}\altaffilmark{1}
and
Herman \textsc{Hensberge}\altaffilmark{3}}
\altaffiltext{1}{\c{C}anakkale Onsekiz Mart University, Physics Department and Ulup{\i}nar Observatory, Terzio\v{g}lu Campus, TR-17020, \c{C}anakkale, Turkey}
\email{bakisv@comu.edu.tr}
\email{ibulut@comu.edu.tr}
\email{bhicran@comu.edu.tr}
\email{demircan@comu.edu.tr}
\altaffiltext{2}{\.{I}stanbul University Science Faculty, Department of Astronomy and Space Sciences, 34119 University-\.{I}stanbul, Turkey}
\email{sbilir@istanbul.edu.tr}
\altaffiltext{3}{Royal Observatory of Belgium, Ringlaan 3, 1180 Brussels, Belgium}
\email{herman@oma.be}


\maketitle

\begin{abstract}
New high resolution spectra of the short period ($P\sim$1.76 days) young detached binary LT CMa are reported for the first time. By combining the results from the analysis of new radial velocity curves and published light curves, we determine values for the masses, radii and temperatures as follows: $M_{\rm 1}=$ 5.59 (0.20) $M_{\rm \odot}$, $R_{\rm 1}=$ 3.56 (0.07) $R_{\odot}$ and  $T_{\rm eff1}=$ 17000 (500) $K$ for the primary and $M_{\rm 2}=$ 3.36 (0.14) $M_{\odot}$, $R_{\rm 2}=$ 2.04 (0.05) $R_{\odot}$ and $T_{\rm eff2}=$ 13140 (800) $K$ for the secondary. Static absorbtion features apart from those coming from the close binary components are detected in the several spectral regions. If these absorbtion features are from a third star, as the light curve solutions support, its radial velocity is measured to be $RV_{\rm 3}=$70(8) km s$^{-1}$.

The orbit of the binary system is proved to be eccentric ($e=$0.059) and thus the apsidal motion exists. The estimated linear advance in longitude of periastron corresponds to an apsidal motion of $U=69\pm$5 yr for the system. The average internal structure constant $\log k_{\rm 2,obs}$=--2.53 of LT CMa is found smaller than its theoretical value of $\log k_{\rm 2,theo}$=--2.22 suggesting the stars would have more central concentration in mass. The photometric distance of LT CMa ($d=$535$\pm$45 pc) is found to be much smaller than the distance of CMa OB1 association (1150 pc) which rules out membership. A comparison with current stellar evolution models for solar metallicity indicates that LT CMa (35 Myr) is much older than the CMa OB1 association (3 Myr), confirming that LT CMa is not a member of CMa OB1. The kinematical and dynamical analysis indicate LT CMa is orbiting the galaxy in a circular orbit and belongs to the young thin-disk population.
\end{abstract}

\section{Introduction}

Detailed analysis of early-type double-lined eclipsing binaries (dEBs) in OB associations yields not only the precise absolute stellar parameters but also information about the association in which they are embedded. The systemic velocity of dEBs, when combined with their distances, allows us to determine the kinematics of the association. The evolutionary stage of the dEBs projects the age of the association, and the chemical composition of the components of dEBs gives the chemical composition of all member stars of the association under the assumption that all member stars are formed in the same medium. This valuable information was successfully obtained for a number of OB associations by applying modern analysis techniques on early-type systems (i.e. among others, $\eta$ Mus (Bak{\i}\c{s} et al., 2007); DW Car (Southworth \& Clausen, 2007); AG Per (Gim\'{e}nez \& Clausen, 1994)). In the frame of a project aimed to investigate early-type dEBs in the vicinity of OB associations, we observed several dEBs in the vicinity of the CMa OB1 association. Some are found to be members and others not.

LT CMa (also HIP 34080, HD 53303) is a relatively bright ($V\sim$7.4 mag), short-period ($P_{\rm orb}$=1.759535 days), eclipsing binary whose light variations were discovered with the {\em Hipparcos} satellite (ESA, 1997). To our knowledge, there is no detailed photometric or spectroscopic study of the system to date. Claria (1974) reported the Johnson colors of LT CMa as $B-V=-0.11$ and $U-B=-0.57$. Otero (2005) comparing the {\em Hipparcos} (ESA, 1997) and ASAS (Pojmanski, 2002) photometric data interpreted an apsidal motion in the system from the shift of the secondary minima. Recently, LT CMa is listed in the catalogue of eclipsing binaries with eccentric orbits by Bulut \& Demircan (2007).

In the present study, new high-resolution ($R\sim$ 40000) spectroscopic observations are used together with all photometric data available from surveys of LT CMa to reveal the apsidal motion rate of the system (section 5) and the absolute stellar parameters of the components (section 6.1). A discussion on the membership of LT CMa to the CMa OB1 association on the basis of evolutionary properties is given in section 6.2. The kinematical properties and the population type of LT CMa are determined in section 6.3. Finally, we calculated the internal structure constants in section 6.4 for a comparison with theoretical predictions.

\section{Spectroscopic Observations and Data Reduction}

LT CMa has been spectroscopically observed using the Coud\'{e}-\'{E}chelle Spectrograph (CES) at the 1.5-m RTT150 telescope of the T\"{U}B\.{I}TAK National Observatory (TUG). Presently, CES gives spectra with a resolving power of $R\sim$40000 and is able to provide 68 spectral orders between 3956-8772 \r{A} in a single CCD frame. The detector is a SAO RAS 1k$\times$1k nitrogen cooled CCD camera. Technical specifications of the CES are described by Bikmaev et al. (2005).

In total, 14 spectra have been collected in five observing runs. The log of the observations is given in table 1, where the HJD of the observation corresponds to the mid-exposure times. For wavelength calibration, Thorium-Argon lamp spectra were taken in the beginning and at the end of each run. Each image was dealt with using IRAF\footnote{IRAF is distributed by the National Optical Astronomy Observatories, which is operated by the Association of Universities for Research in Astronomy, Inc. (AURA) under cooperative agreement with the National Science Foundation.} according to normal procedures of bias substraction, flat-field division, background-light subtraction and wavelength calibration. For the orders where very broad stellar lines such as H$_\alpha$ and He I (4471 \AA) are near the order edges, the continuum correction for these orders has been performed by means of dividing the order by the normalized continuum function of the previous or next order where no stellar lines exist.

\begin{table}
\begin{center}
\caption{Journal of spectroscopic observations. Signal-to-noise (S/N) ratio refers to the continuum near 5700\,\AA. \label{table1}}
\begin{tabular}{cccccr}
\hline
ID  &    HJD     &   S/N      & \multicolumn{1}{c}{Exposure Time} \\
    & (-2450000) &            & \multicolumn{1}{c}{(s)}   \\
\hline
1  & 4786.50846 &  50   & 3200 \\
2  & 4787.44775 &  30   & 2500 \\
3  & 4787.51254 &  30   & 2500 \\
4  & 4787.54492 &  50   & 3000 \\
5  & 4839.42717 &  50   & 2400 \\
6  & 4839.45624 &  50   & 2400 \\
7  & 4839.48497 &  50   & 2400 \\
8  & 4839.56654 &  30   & 2400 \\
9  & 4840.29345 &  40   & 3600 \\
10 & 4840.39241 &  40   & 3600 \\
11 & 4840.48616 &  35   & 3600 \\
12 & 4840.52931 &  20   & 3600 \\
13 & 4841.34322 &  55   & 2400 \\
14 & 4841.37807 &  65   & 2400 \\
\hline \\
\end{tabular}
\end{center}
\end{table}

\section{Radial Velocities and Spectroscopic Orbit}

The components of the close binary show broad lines. In addition, a careful inspection of the spectra at different orbital phases reveals at several locations weak, static absorption features that do not participate in the Doppler motions. These static features include H$_\alpha$, H$_\delta$ and Mg II (4481 \AA) and may be due to a third star that was not detected in the He I lines. Since the latter are among the strongest absorption lines in B-type stars, the static features may be associated with a cooler third star. The composite spectra of the close binary components and the static absorption features at three orbital phases are shown in figure 1.

The radial velocity ($RV$) of the static absorption features is  71.2 (7.1) km s$^{-1}$ from the H$_\alpha$ line and 68.2 (7.7) km s$^{-1}$ from the Mg II (4481\,\AA) line. Errors given in brackets are the standard deviations of the measurements which were made by means of fitting Gaussian to the central part of the line. Finally, the mean $RV$ of is about 70 (8) km s$^{-1}$.

The strong blending of the broad Balmer series hydrogen lines of the (three) components makes them useless for $RV$ measurements. We selected the spectral order 53 for $RV$ study, where relatively strong He I (4471\,\AA) and Mg II (4481\,\AA) lines reside. We attempted to measure the Doppler shifts of individual spectral lines by fitting a Gaussian to the central part of individual lines. However, in some spectra no reliable measurement is possible, either due to the lower signal-to-noise ratio or to the smaller Doppler shifts, or to a combination of both. At these phases, radial velocities were not measured and they are left blank in table 2, where measured $RVs$ are presented.

\begin{table}
\begin{center}
\caption{Radial velocities of LT CMa. Ephemeris given in table 3 was used as zero epoch for phase calculation. Values in brackets are the $O-C$ residuals from the theoretical Keplerian orbital fitting.\label{table2}}
\begin{tabular}{cccc}
        &           &             &            \\
\hline
Time    &   Phase   &  $RV_{\rm 1}$   &   $RV_{\rm 2}$ \\
HJD     &   $\phi$  &(km s$^{-1}$)&  (km s$^{-1}$) \\
\hline
2454786.50846 & 0.184 & -108.3 (4.4) &  242.5 (2.9) \\
2454787.44775 & 0.718 & -            & -       \\
2454787.51254 & 0.755 &  153.2 (7.1) & -172.1 (10.0) \\
2454787.54492 & 0.773 &  138.4 (1.1) & -180.9 (10.7) \\
2454839.42717 & 0.259 & -109.9 (5.0) &  267.1 (8.4)  \\
2454839.45624 & 0.276 & -116.0 (7.6) &  242.9 (10.6) \\
2454839.48497 & 0.292 & -115.4 (5.4) &  251.5 (0.6) \\
2454839.56654 & 0.339 & -            & -       \\
2454840.29345 & 0.752 & -            & -       \\
2454840.39241 & 0.808 &  134.8 (7.9) & -165.1 (6.7) \\
2454840.48616 & 0.861 & -            & -150.6 (9.9) \\
2454840.52931 & 0.886 & -            & -       \\
2454841.34322 & 0.348 & -            & -       \\
2454841.37807 & 0.368 & -81.8  (6.4) & -       \\
\hline \\ \\ \\
\end{tabular}
\end{center}
\end{table}

\begin{figure}
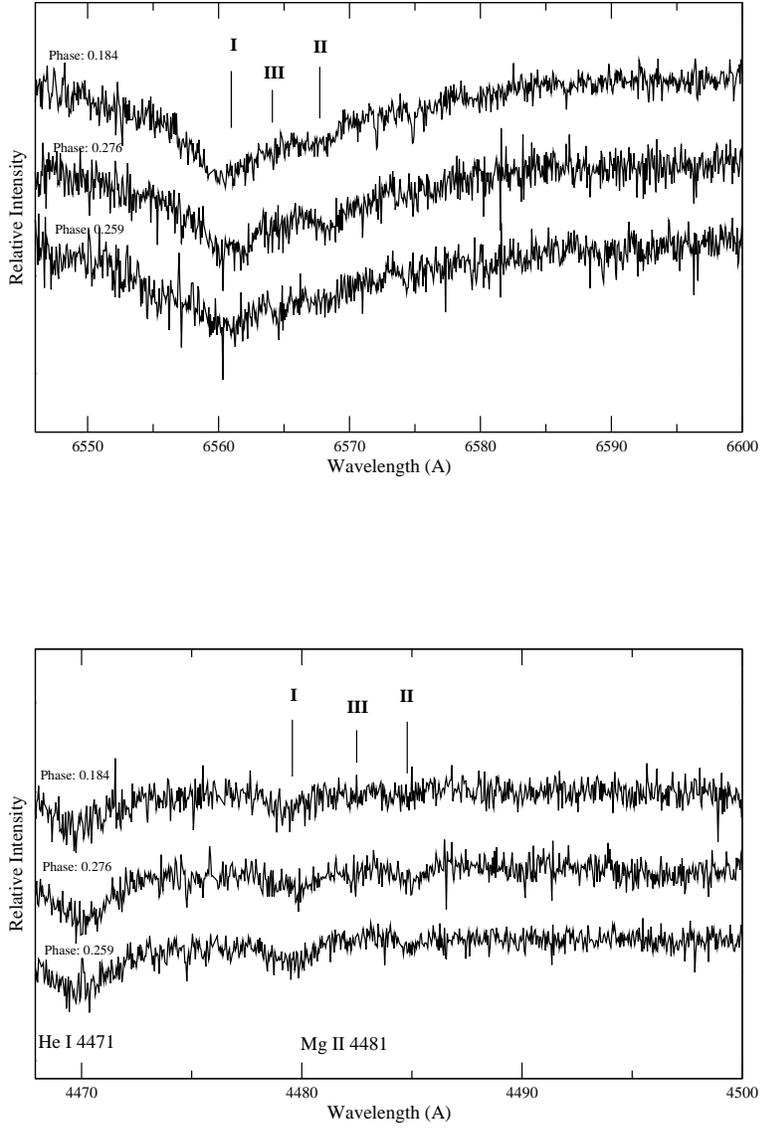

\begin{center}
\begin{tabular}{c}
\FigureFile(100mm,100mm){figure1a.eps} \\ \\ \\ \\
\FigureFile(100mm,100mm){figure1b.eps}
\end{tabular}
\caption{H$_\alpha$ (top) and Mg II (4481\,\AA) (bottom) lines of the components at three orbital phases. In each panel, I, II and III represent the lines of the primary, the secondary and the third stars, respectively.} \label{fig1}
\end{center}
\end{figure}

Keplerian spectroscopic orbits have been fitted to the $RVs$ by means of a differential corrections technique. The ephemeris for primary and secondary minima were derived by Otero (2005) as given in Eqs. 1 and 2. We adopted the orbital period of LT CMa from the primary minimum ephemeris (Eq. 1) and fixed it during the orbital solutions. Other parameters such as velocity semi-amplitudes of the components ($K_{\rm 1, 2}$), systemic velocity ($RV_\gamma$), longitude of periastron ($\omega$) and time of periastron passage ($T_{\rm 0}$) were converged. The orbital eccentricity of LT CMa is close to zero (see table 4). In view of the low number of available spectra and the lack of spectra at quadrature phases, $e$=0.059 was kept fixed at the value obtained from light curve analysis (see table 5), rather than to let it converge to a spurious value. The orbital solution is presented in figure 2 and table 3.

\begin{eqnarray}
{\rm Min~I (HJD)}  = 2448388.537 + 1.759535 \times E, \\
{\rm Min~II (HJD)} = 2448389.445 + 1.759503 \times E.
\end{eqnarray}

\begin{figure}
\begin{center}
\begin{tabular}{c}
\FigureFile(100mm,100mm){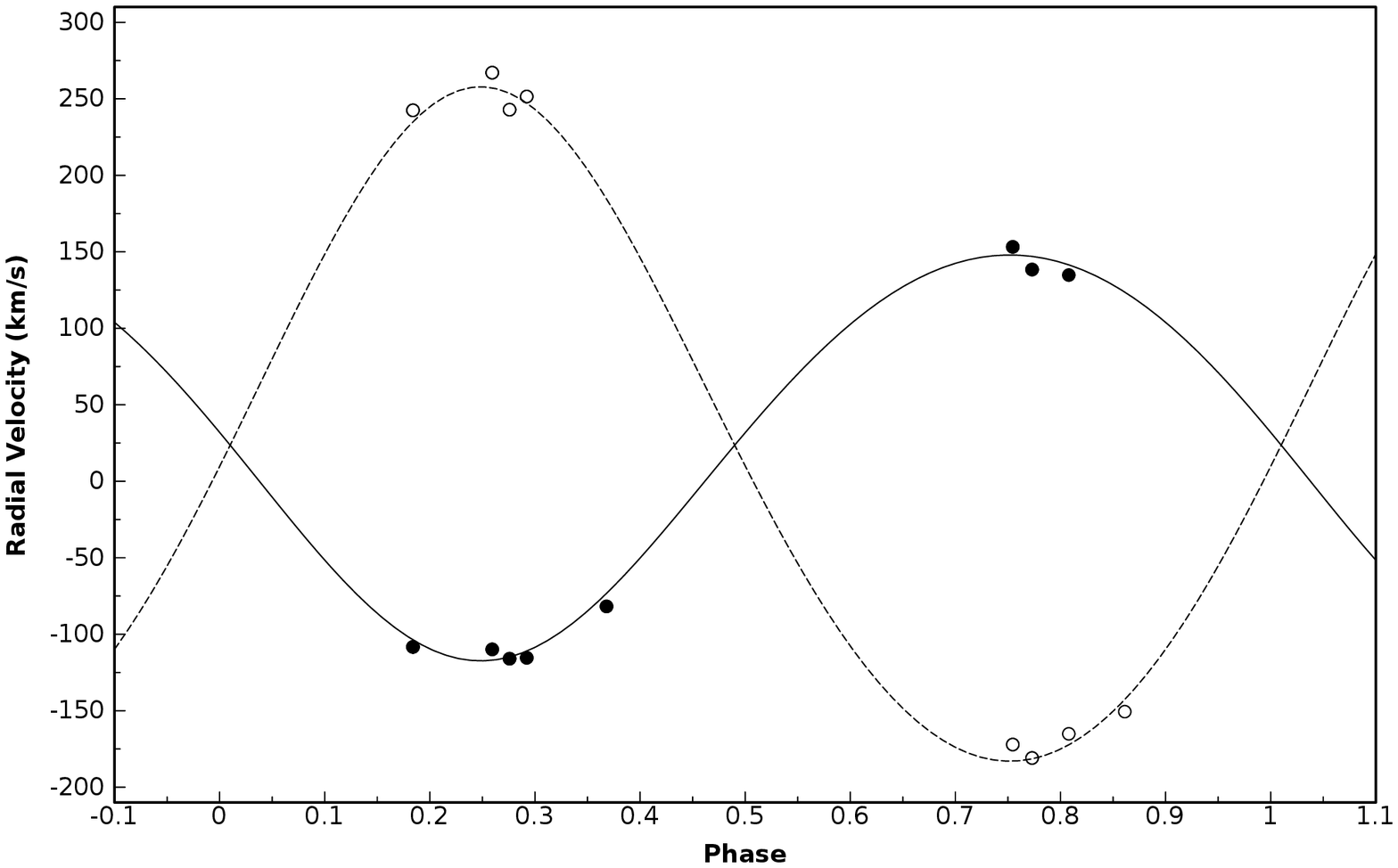}
\end{tabular}
\caption{Theoretical curve fitting to the $RV$ data. The $RVs$ for primary and secondary stars are shown with filled and empty circles, respectively.} \label{fig2}
\end{center}
\end{figure}

\begin{table}
\begin{center}
\caption{The adopted spectroscopic orbital parameters of LT CMa. \label{table3}}
\begin{tabular}{lc}  \hline\hline
Parameter                    & Value                  \\
\hline
$P$ (days)                   & 1.759535 (fixed)       \\
$T_{\rm 0}$(HJD-2454000)         & 786.1848 $\pm$ 0.0124  \\
$e$                          & 0.059 (fixed)          \\
$\omega_{\rm sp}$ (rad)          &  3.05 $\pm$ 0.36       \\
$K_{\rm 1}$ (kms $^{-1}$)        & 132.6 $\pm$ 2.9        \\
$K_{\rm 2}$ (kms $^{-1}$)        & 220.4 $\pm$ 3.1        \\
$q$                          & 0.602 $\pm$ 0.020      \\
$RV_{\rm \gamma}$  (kms$^{-1}$)  &  23.6 $\pm$ 2.0        \\
$m_{\rm 1} \sin^{3}i$ ($M_{\odot}$)& 4.977 $\pm$ 0.070    \\
$m_{\rm 2} \sin^{3}i$ ($M_{\odot}$)& 2.995 $\pm$ 0.065    \\
$a\sin i$ ($R_{\odot}$)      & 12.26 $\pm$ 0.20       \\
RMS (kms $^{-1}$)            & 7.44                   \\
\hline
\end{tabular}
\end{center}
\end{table}

For a check of our spectroscopic orbital solution, the spectral disentangling technique was used in the region of He I (4471 \,\AA\,) and Mg II (4481\,\AA\,) lines. The spectral disentangling technique has the advantage to find the spectroscopic-orbit elements and the component spectra simultaneously without requiring the intermediate step of radial velocity measurements on stellar lines (cf. Hadrava, 1995). It yields, in our case, perfectly consistent results with the $RVs$ obtained from Gaussian fitting. However, spurious patterns in the reconstructed component spectra, especially at the edges, indicate the necessity to combine the spectral orders in an accurate way, such that the edges of the reconstructed regions are in the continuum, and, possibly, the addition of the third component. The present data sets is also non-optimal in the sense that it covers the extreme Doppler shifts, but not their whole range, and that the eclipses are not covered. The different line-dilution in mid-eclipse spectra are required to obtain a good solution of the low-frequency Fourier components of the component spectra. Summarizing, a successful exploitation of the disentanglement techniques will be undertaken when the data set is appropriately extended.

\section{Analysis of Light Curves}

\subsection{Binary Model and Input Parameters}

Analysis of the light curves was achieved on the basis of three different photometric data sets ($H_{p}$-band data of {\em Hipparcos} (ESA, 1997), $V$-band data of ASAS (Pojmanski, 2002) and {\em INTEGRAL} surveys). {\em Hipparcos} and {\em INTEGRAL} observations have in total 68 and 480 photometric measurements, respectively. The usefulness of the {\em Hipparcos} data is limited by lack of observations in the secondary minimum and the usefulness of the {\em INTEGRAL} data is limited by insufficient sensitivity. These data sets were used only for the determination of periastron longitude. The 315 photometric measurements of ASAS, including eclipses, were used for the determination of light-curve elements. All available photometric data are phased and re-scaled in figure 3.

\begin{figure*}
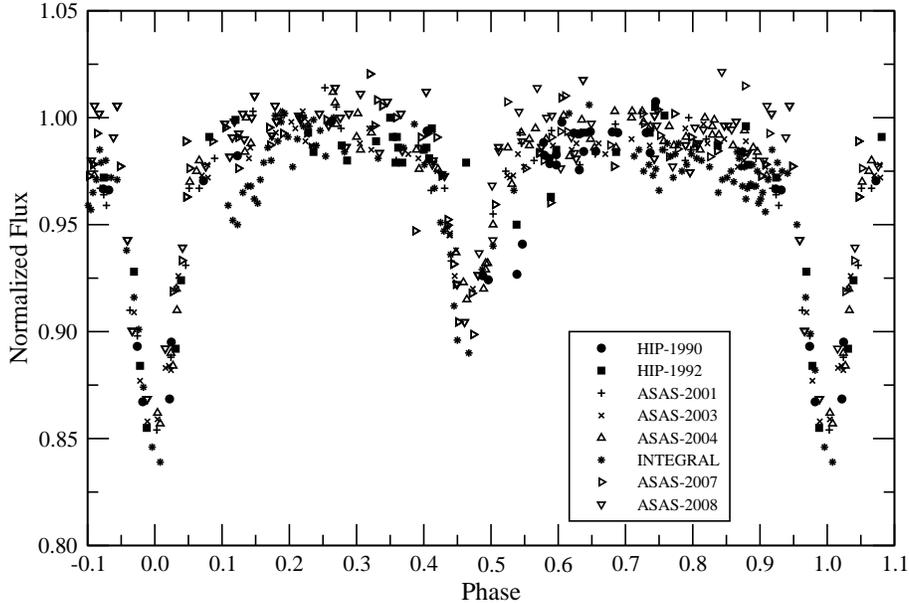

\begin{center}
\vspace{2cm}
      \FigureFile(120mm,120mm){figure3.eps}
\caption{\textit{Hipparcos}, ASAS and \textit{INTEGRAL} light curves of LT CMa. Observing seasons are indicated with survey name using different symbols.}
\label{fig3}
\end{center}
\end{figure*}

Otero (2005), using {\em Hipparcos} and ASAS photometric data, commented the existence of apsidal motion in the system without mentioning its rate. Apsidal motion may lead to wrong light curve elements when photometric data is not carefully studied. Especially, when the data is spread over years, one should pay attention to analyze subsets separately, for example each observing season separately, in order to reduce the influence of apsidal motion in a specific set to an insignificant level (cf. Bak{\i}\c{s} et al. 2008). In the present paper, we use 5 subsets of ASAS data over a time base of 8 years (2001 -- 2009). The 2003 and 2004 data covering the whole orbital cycle, were found to be more suitable for a reliable light-curve analysis. We discarded subsets of 2005, 2006, 2009 due to either low number of data or relatively large scatter. Except for the 2007 data sets of ASAS and the {\em INTEGRAL}, which were analyzed together, we performed a separate analysis for each ASAS and {\em Hipparcos} data set using the 2003 version of the Wilson \& Devinney (WD; Wilson \& Devinney 1971; Wilson 1994) program.

The effective temperature of the primary component must be determined for the light-curve modeling. The $Q$-method of Johnson \& Morgan (1953) has been used to estimate unreddened colours and interstellar reddening. Using $U-B=-$0.57(0.01) and $B-V=-$0.11(0.009) colours given by Claria (1974), $Q=-$0.49(0.02) and the unreddened colour $(B-V)_{\rm 0}=-$0.17(0.03).

Consequently the colour excess $E(B-V)=$0.06(0.03), $(U-B)_{0}=-$0.61(0.03) and the visual extinction towards LT CMa $A_{\rm V}=$0.19(0.03) mag. These unreddened colours correspond to an interpolated spectral type of B4.5V (Fitzgerald, 1970), which corresponds to a temperature of 16200 K according to the calibration tables of Siess, Forestini \& Dougados (1997). However, it should be noticed that these colours and the corresponding temperature are obtained from the combined light of components including the third companion, resulting in a slightly redder colour and lower temperature than the intrinsic colour of the primary component. The light curve and the spectral lines show that the light of the primary star dominates largely. Therefore, it is reasonable to adopt a temperature of 17000 K which is also consistent with the spectral type of the primary (see section 6.1).

\subsection{Determination of the Photometric Elements}

With the analysis strategy of the light curves and the input parameters determined in section 4.1, the solution of the light curves has been performed. The temperature of the primary was fixed at $T_{\rm eff1}$=17000 K and the temperature of the secondary was left to converge to $T_{\rm eff2}$. The third light contribution to the total flux of the system was set initially 0.1, based on the line strengths of the static component in H$_\alpha$ and Mg II (4481\,\AA\,). Gravity darkening exponents $g_{\rm 1}=g_{\rm 2}=1$ and bolometric albedos $A_{\rm 1}=A_{\rm 2}=1$ were set for radiative envelopes (von Zeipel, 1924). Logarithmic limb-darkening law was used and limb-darkening coefficients were taken from van Hamme (1993). The surface potentials ($\Omega_{\rm 1,2}$), light factors of the components ($l_{\rm 1,2}$), orbital inclination ($i$), eccentricity ($e$) and longitude of periastron ($\omega$) were the adjustable parameters during the light curve modeling. The initial guess of the longitude of periastron ($\omega$) was taken from the spectroscopic solution in section 3. Mass ratio was fixed at the spectroscopic value of $q=0.602$.

The solutions of $V$-band ASAS light curves of 2003 and 2004 years converged very rapidly, and had the smallest residuals. The third light contribution did not deviate significantly from the initial guess. Input parameters were then altered to check the consistency and uniqueness of the solution. These new input parameters were converged to the parameters of the first solution, which shows the consistency of solutions listed in table 4.

Using the light curve elements from ASAS data sets (2003, 2004) as fixed parameters, {\em Hipparcos}, {\em INTEGRAL} and ASAS (2001, 2007, 2008) light curves were analyzed to determine the longitude of the periastron. Since the {\em Hipparcos} data are also spread over three years, the {\em Hipparcos} measurements were also divided into two data sets for the years 1990 and 1992. During the fitting to the observational data sets (1990, 1992, 2001, 2003, 2004, 2007 and 2008), the longitude of the periastron and the light factors of the components were the parameters adjusted. The analysis of each light curve set yielded different values of $\omega$ (see section 5) which clearly showed a time dependent apsidal motion. Apsidal motion is investigated in section 5 in more detail.

Adopted light curve solutions for each data sets of LT CMa are shown in figure 4. The light curve elements obtained from these data sets are given in table 4. For clarity, we give the adopted light curve elements in table 5.

\begin{table*}\small
\begin{center}
\caption{Light curve fitting results for LT CMa. The parameters converged during the fittings are shown with their errors. The parameters without errors are those adopted from the solutions of 2003 and 2004 light curves (last two columns). The goodness of the fits to each data sets are given in the bottom of the table as $\chi^{2}$ together with the number of observations (NOBS) in the respective data set. Note results for 2007 are from two data sets ASAS and {\em INTEGRAL}. NOBS and $\chi^{2}$ are given in brackets accordingly.}\label{table4}
\begin{tabular}{lccccccc}
\noalign{\smallskip} \hline \hline \noalign{\smallskip}
Parameters      & $Hp_{\rm 1989}$ &$Hp_{\rm 1991}$  &$V_{\rm 2001}$&$V_{\rm 2007}$ &$V_{\rm 2008}$&$V_{\rm 2003}$       &$V_{\rm 2004}$       \\
\noalign{\smallskip} \hline \noalign{\smallskip}
$T_{\rm h}$ (K)     & 17000           & 17000           & 17000        & 17000         & 17000        & 17000           & 17000           \\
$T_{\rm c}$ (K)     & 13140           & 13140           & 13140        & 13140         & 13140        & 13140$\pm$210   & 13135$\pm$315   \\
$q~(M_{\rm 2}/M_{\rm 1})$   & 0.602           & 0.602           & 0.602        & 0.602         & 0.602        & 0.602           & 0.602           \\
$e$             & 0.059           & 0.059           & 0.059        & 0.059         & 0.059        & 0.058$\pm$0.003 & 0.059$\pm$0.006 \\
$w~(rad)$       & 1.23$\pm$0.09   & 1.39$\pm$0.09   & 2.40$\pm$0.11& 2.80$\pm$0.11 & 2.60$\pm$0.16& 2.39$\pm$0.08   & 2.55$\pm$0.16   \\
$l_{\rm 1}$         & 0.757$\pm$0.011 & 0.744$\pm$0.011 & 0.737        &0.737          &0.737         & 0.732$\pm$0.008 & 0.743$\pm$0.008 \\
$l_{\rm 2}$         & 0.163$\pm$0.015 & 0.161$\pm$0.015 & 0.162        &0.162          &0.162         & 0.164$\pm$0.012 & 0.159$\pm$0.012 \\
$l_{\rm 3}$         & 0.080$\pm$0.020 & 0.095$\pm$0.020 & 0.102        &0.102          &0.102         & 0.104$\pm$0.021 & 0.098$\pm$0.020 \\
$\Omega_{\rm 1}$    & 4.25            & 4.25            & 4.25         & 4.25          & 4.25         & 4.204$\pm$0.041 & 4.300$\pm$0.055 \\
$\Omega_{\rm 2}$    & 5.06            & 5.06            & 5.06         & 5.06          & 5.06         & 5.012$\pm$0.062 & 5.133$\pm$0.081 \\
$r_{\rm 1}$ (pole)  & 0.274           & 0.274           & 0.274        & 0.274         & 0.274        & 0.277$\pm$0.003 & 0.272$\pm$0.004 \\
$r_{\rm 1}$ (point) & 0.289           & 0.289           & 0.289        & 0.289         & 0.289        & 0.292$\pm$0.003 & 0.286$\pm$0.004 \\
$r_{\rm 1}$ (side)  & 0.280           & 0.280           & 0.280        & 0.280         & 0.280        & 0.282$\pm$0.003 & 0.277$\pm$0.004 \\
$r_{\rm 1}$ (back)  & 0.286           & 0.286           & 0.286        & 0.286         & 0.286        & 0.289$\pm$0.003 & 0.283$\pm$0.004 \\
$r_{\rm 1}$ (volume)& 0.280           & 0.280           & 0.280        & 0.280         & 0.280        & 0.283$\pm$0.003 & 0.277$\pm$0.004 \\
$r_{\rm 2}$ (pole)  & 0.157           & 0.157           & 0.157        & 0.157         & 0.157        & 0.159$\pm$0.002 & 0.155$\pm$0.003 \\
$r_{\rm 2}$ (point) & 0.159           & 0.159           & 0.159        & 0.159         & 0.159        & 0.162$\pm$0.002 & 0.157$\pm$0.003 \\
$r_{\rm 2}$ (side)  & 0.157           & 0.157           & 0.157        & 0.157         & 0.157        & 0.160$\pm$0.002 & 0.155$\pm$0.003 \\
$r_{\rm 2}$ (back)  & 0.159           & 0.159           & 0.159        & 0.159         & 0.159        & 0.162$\pm$0.002 & 0.157$\pm$0.003 \\
$r_{\rm 2}$ (volume)& 0.158           & 0.158           & 0.158        & 0.158         & 0.158        & 0.160$\pm$0.002 & 0.156$\pm$0.003 \\
$i$ ($^\circ$)  & 74.2            & 74.2            & 74.2         & 74.2          & 74.2         & 73.9$\pm$0.3    & 74.5$\pm$0.4    \\
NOBS            &32               & 36              & 39           & 480, 64       & 78           &  66             & 68              \\
$\chi^{2}$      &0.001527         & 0.002589        & 0.003311     & 0.147187, 0.011011 & 0.011290&  0.001600       & 0.002633        \\
\noalign{\smallskip} \hline \noalign{\smallskip}
\end{tabular}
\end{center}
\end{table*}

\begin{table}
\begin{center}
\caption{Adopted V-band light curve elements for LT CMa. $\Omega_{\rm cr}$ stands for the potential of the inner critical Roche lobe.}\label{table5}
\begin{tabular}{lcc}
\hline\hline
Parameter        &     Primary        &     Secondary       \\
\hline
$T\,(K)$         & 17000              &     13140$\pm$315   \\
$q\,(M_{\rm 2}/M_{\rm 1})$   & \multicolumn{2}{c}{0.602}                \\
$i\,(^{\circ})$  & \multicolumn{2}{c}{74.2$\pm$0.3}         \\
$e$              & \multicolumn{2}{c}{0.059$\pm$0.006}      \\
$l\,(l_{\rm 1,2}/l_{\rm Total})$& 0.737$\pm$0.008 & 0.162$\pm$0.012 \\
$l\,(l_{\rm 3}/l_{\rm Total})$& \multicolumn{2}{c}{0.102$\pm$0.020} \\
$\Omega$         & 4.25$\pm$0.04      & 5.06$\pm$0.06       \\
$\Omega_{\rm cr}$& \multicolumn{2}{c}{3.22} \\
$r$              & 0.280$\pm$0.004    & 0.160$\pm$0.003     \\
\hline
\end{tabular}
\end{center}
\end{table}

\begin{figure*}
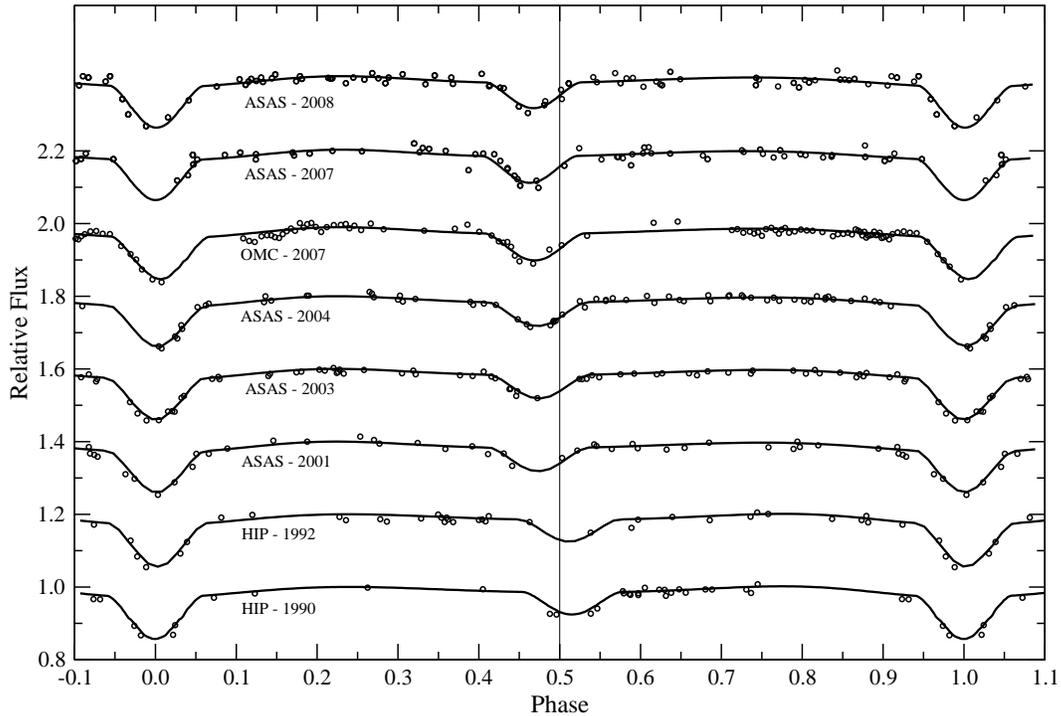

\begin{center}
\begin{tabular}{c}
      \FigureFile(140mm,140mm){figure4.eps} \\
\end{tabular}
\caption{Photometric solutions of $V$ and $H_{p}$-band light curves.} \label{fig4}
\end{center}
\end{figure*}

\section{Apsidal Motion}

The orbit of LT CMa has been determined to be slightly eccentric ($e$ = 0.059) from the solutions of the light curves in section 4. This indicates that LT CMa is a system showing apsidal motion. However, due to the small number and unfavorable distribution of the times of minima, it is not possible to determine the apsidal motion period by the $O-C$ variations. Therefore, determination of the apsidal motion rate has been performed by means of linear least-squares fitting to the longitude of periastron ($\omega$) values which were obtained from nine different data sets (two \textit{Hipparcos} light curves, five ASAS light curves, one Integral light curve and the spectroscopic data). The best linear fit ($\omega$ = $\omega_{\rm 0}$ + $\dot{\rm \omega}$ $T$) to the set of longitude of periastron values is presented in figure 5. The data are consistent with a linear advance in longitude of periastron, $\dot{\rm \omega}$=0.0252 $\pm$ 0.0020 deg cycle$^{-1}$. This apsidal motion rate corresponds to an apsidal motion period of $U$ = 69 $\pm$ 5 yr which is a relatively short apsidal period compared to the eccentric binary systems given by (Bulut \& Demircan 2007).

\begin{figure}
\begin{center}
\vspace{1cm}
\begin{tabular}{c}
      \FigureFile(120mm,120mm){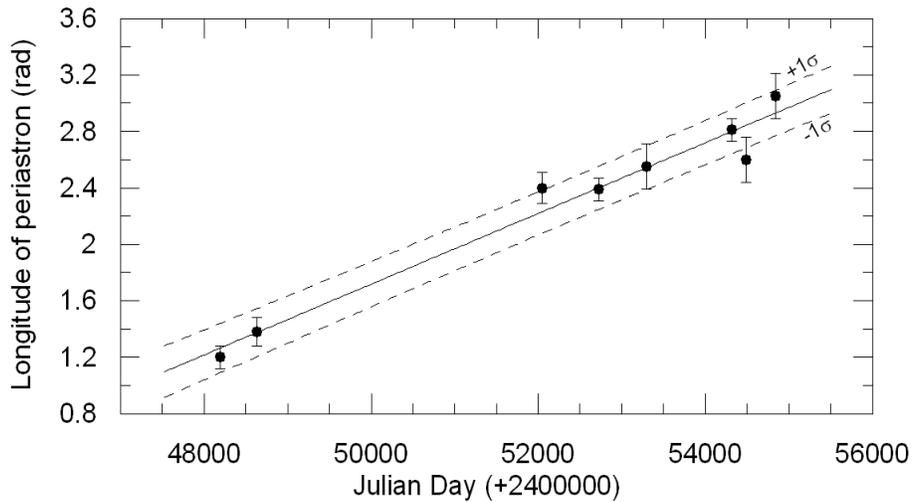} \\ \\ \\
\end{tabular}
\caption{Apsidal motion calculation through a linear regression to longitude of periastron ($\omega$) values obtained from the analysis of eight light curves and one $RV$ curve. Least squares fit to the $\omega$ values with their uncertainties yielded the slope of the straight line as 0.00025$\pm$0.00002 rad d$^{-1}$. The $\pm$1-$\sigma$ levels is also shown with dashed lines.}
\label{fig5}
\end{center}
\end{figure}

\section{Discussion}

\subsection{Absolute dimensions and distance of LT CMa}

The fundamental astrophysical parameters of LT CMa, which were derived from spectroscopic results in table 3 and from the light-curve results in table 4, are summarized in table 6. The temperature $T_{\rm eff1} =$ 17000 K, mass $M_{\rm 1}=5.59M_{\odot}$ and radii $R_{\rm 1}=3.56R_{\odot}$ of the primary component correspond to a spectral type B4V. The temperature $T_{\rm eff2} =$ 13140 K of the secondary star implies a B6--B7 spectral type ZAMS star which is also very consistent with its mass of $M_{\rm 2}=3.36M_{\odot}$ and radius of $R_{\rm 2}=2.04R_{\odot}$ (i.e. Strai\v{z}ys \& Kuriliene 1981).

The unreddened Johnson $V$-magnitude (Clarie 1974) of LT CMa, when combined with the light contributions as derived from the light-curve analysis, including the distant companion, yields the intrinsic $V$-magnitudes $m_{\rm V1}=$7$^m$.55, $m_{\rm V2}=$9$^m$.19 and $m_{\rm V3}=$9$^m$.72, respectively. Using $M_{\rm v}=$4$^m$.75 as the absolute visual magnitude of the Sun and bolometric corrections $BC_{\rm 1}=$--1.58 and $BC_{\rm 2}=$--0.93 mag for the primary and the secondary, from Straizys \& Kuriliene (1981), bolometric and absolute visual magnitudes of the close binary components are derived (see table 6). The visual magnitude and distance modulus indicate a photometric distance of 535 $\pm$ 45 pc to the LT CMa, which is smaller than the distance (826$\pm$326 pc) derived for LT CMa by van Leeuwen (2007) from the re-analysis of raw Hipparcos data, but not inconsistent with it in view of the large error bar on the Hipparcos distance. In the original Hipparcos catalogue, the distance to LT CMa is given to be 591$\pm$386 pc. Among these measurements, the photometrically determined distance of LT CMa comes out to be more precise and reliable.

The distance of the CMa OB1 association (1150 pc) (Claria 1974) is much greater than the distance of LT CMa (535 $\pm$ 45 pc), which weakens the probability that LT CMa belongs to the association.

The projected rotational velocities of the components were derived to be $V_{\rm rot1}$$\times$$\sin(i)$ = 105(10) km s$^{-1}$ and
$V_{\rm rot2}$$\times$$\sin(i)$ = 65(10) km s$^{-1}$ by fitting synthetic spectra using Kurucz (1993) atmosphere models and the ATLAS9 code (Kurucz 1993). Using the orbital inclination ($i$=74.2 deg) of LT CMa, the observed rotational velocities of the components are found to be $V_{\rm rot1}$ = 109(10) km s$^{-1}$ and $V_{\rm rot2}$ = 67(10) km s$^{-1}$ which are in agreement with the synchronous rotational velocities within the uncertainties.

Although the light curve solution supports an extra light in the system, the assumption of the third star is still preliminary due to the low number of high S/N spectra used in present study. Thus, it is desired in the future study to obtain high S/N spectra evenly distributed over the whole orbital cycle in order to reveal the identity of additional features in the composite spectrum.

\begin{table*}
\small \caption{Close binary stellar parameters of LT~CMa. Errors of parameters are given in parenthesis.} \label{table6}
\begin{tabular}{lccc}\hline\hline
Parameter                          & Symbol  & Primary & Secondary                    \\
\hline
Spectral type                      & Sp      & B4 V & B6.5 V                        \\
Mass ($M_\odot$)                   & \emph{M}       & 5.59(0.20) & 3.36(0.14)         \\
Radius ($R_\odot$)                 & \emph{R}       & 3.56(0.07) & 2.04(0.05)         \\
Separation ($R_\odot$)             & \emph{a}       & \multicolumn{2}{c}{12.7(0.2)}   \\
Orbital period (days)              & \emph{P}       & \multicolumn{2}{c}{1.759535}   \\
Orbital inclination ($^{\circ}$)   & \emph{i}       & \multicolumn{2}{c}{74.2(0.3)}   \\
Mass ratio                         & \emph{q}       & \multicolumn{2}{c}{0.602(0.020)}\\
Eccentricity                       & \emph{e}       & \multicolumn{2}{c}{0.059(0.006)}\\
Surface gravity (cgs)              & $\log g$       & 4.081(0.014)& 4.347(0.019)      \\
Integrated visual magnitude (mag)  & \emph{V}       &  7.42(0.01)   \\
Integrated colour index (mag)      & $B-V$         &  -0.11(0.009) \\
Colour excess (mag)                &$E(B-V)$&\multicolumn{2}{c}{0.06(0.03)}\\
Visual absorption (mag)            & $A_{\rm V}$ &\multicolumn{2}{c}{0.19(0.03)}\\
Intrinsic colour index (mag)       & $(B-V)_{\rm 0}$&\multicolumn{2}{c}{-0.17(0.03)}\\
Temperature (K)                    & $T_{\rm eff}$ & 17000(500) & 13140(800) \\
Luminosity ($L_\odot$)             & $\log$ \emph{L}& 2.98(0.03) & 2.05(0.07)\\
Bolometric magnitude (mag)         &$M_{\rm bol}$& -2.70(0.05) & -0.36(0.08)    \\
Absolute visual magnitude (mag)    &$M_{\rm v}$  & -1.12(0.05) & 0.57(0.07)    \\
Bolometric correction (mag)        &\emph{BC}& -1.58      & -0.93          \\
Velocity amplitudes (km s$^{-1}$)  &$K_{\rm 1,2}$& 132.6(2.9) & 220.4(3.1)    \\
Systemic velocity (km s$^{-1}$)    &$RV_{\gamma}$ & \multicolumn{2}{c}{23.6(2.0)} \\
Computed synchronization velocities (km s$^{-1}$)& $V_{\rm synch}$ & 102 & 59 \\
Observed rotational velocities (km s$^{-1}$) & $V_{\rm rot}$ & 109(10) & 67(10) \\
Distance (pc)                      &\emph{d} & \multicolumn{2}{c}{535(45)} \\
Proper motion (mas yr$^{-1}$)      &$\mu_\alpha cos\delta$, $\mu_\delta$ & \multicolumn{2}{c}{-6.82(0.83), 3.53(0.71)*} \\
Space velocities (km s$^{-1}$)     & $u, v, w$  & \multicolumn{2}{c}{-27.46(2.12), -5.21(2.15), -12.45(2.24)}\\
\hline
* from {\em Hipparcos} catalogue (van Leeuwen 2007).
\end{tabular}
\end{table*}

\subsection{Evolutionary stage and age of LT CMa}

We have investigated the evolutionary status of LT CMa in the plane of $\log T_{\rm eff}$ - $\log g$ and Mass - $\log T_{\rm eff}$ (figure 5) using the latest evolutionary models and isochrones of Girardi et al. (2000), which include mass loss and moderate overshooting. Assuming a solar metal content, we prepared a set of isochrones corresponding to $Y=0.28$ and $Z=0.02$. From figure 5a, it can be interpreted that both component are consistent with the evolutionary tracks calculated for their masses. The primary star is evolved from the ZAMS, whereas the secondary is still on the ZAMS. The isochrones of 30 Myr and 40 Myr shown in figure 5a and figure 5b implied a mean age of 35(5) Myr for the system. The age of LT CMa derived here is much greater than that derived by Claria (1974) for CMa OB1 (3 Myr), which rules out membership, when combined with the distance difference of LT CMa and the association.

\begin{figure}
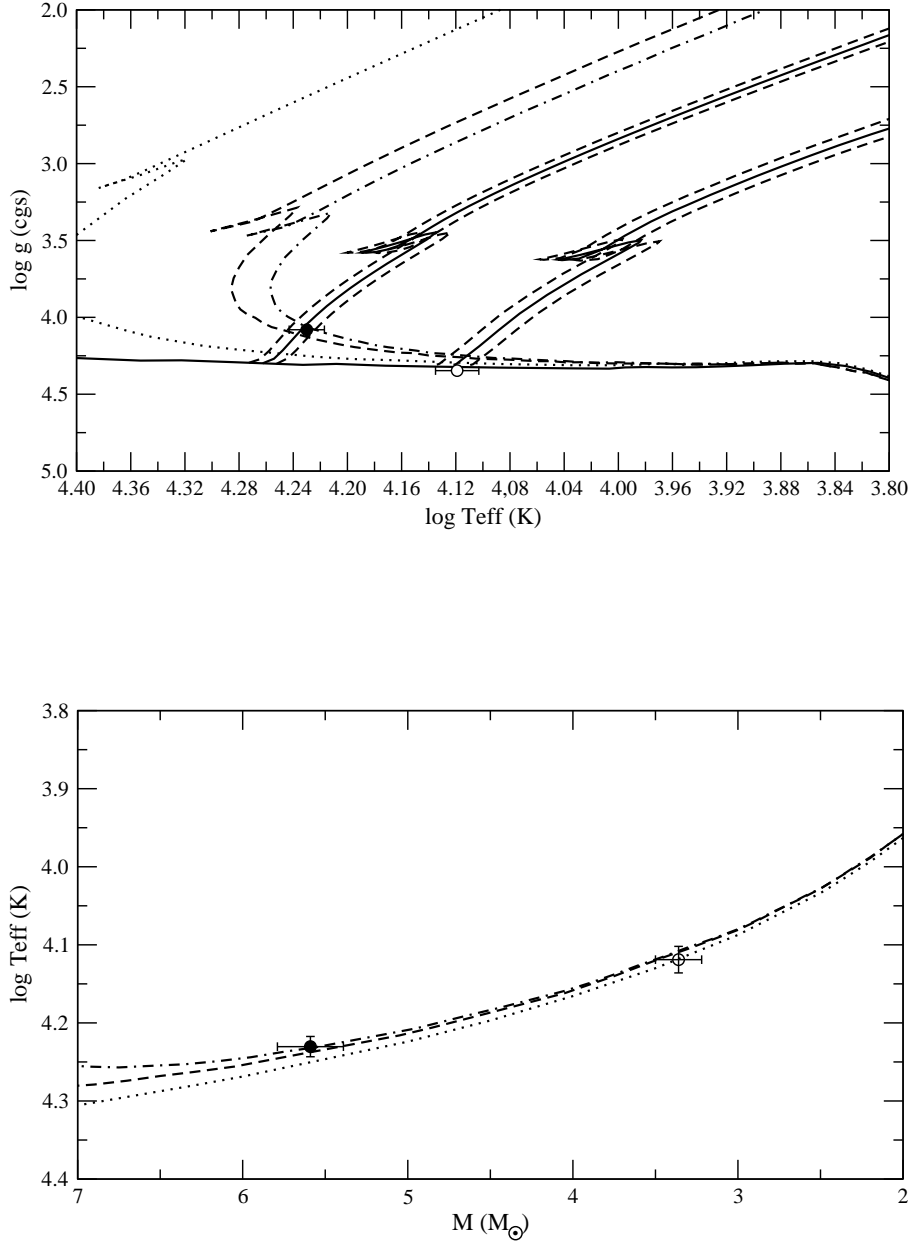

\begin{center}
\vspace{1cm}
\begin{tabular}{c}
\FigureFile(120mm,120mm){figure6a.eps} \\ \\ \\ \\
\FigureFile(120mm,120mm){figure6b.eps} \\ \\
\end{tabular}
\caption{Locations of the components stars in the plane of $\log T_{\rm eff}$ - $\log g$ (upper panel) and in the Mass - $\log T_{\rm eff}$ (lower panel). Primary and secondary stars are shown with filled and empty circles, respectively, together with error bars. Isochrones are for 10 Myr (dotted line), 30 Myr (dash line) and 40 Myr (dotted dash line). Evolutionary tracks for the primary and the secondary stars are presented with solid lines where the dashed lines around evolutionary tracks are due to the uncertainties in masses of the components. The horizontal solid line is for ZAMS.} \label{fig6}
\end{center}
\end{figure}

\subsection{Kinematical analysis and population type of LT CMa}

To study the kinematical properties of LT CMa, we used the system's centre-of-mass velocity, distance and proper motion values, which are given in table 6. The proper motion data were taken from the newly reduced Hipparcos catalogue (van Leeuwen 2007), whereas the center-of-mass velocity and distance are obtained in the present study. The system's space velocity was calculated using the Johnson \& Soderblom's (1987) algorithm. The $u$, $v$ and $w$ space velocity components and their errors are listed in table 6. To obtain the space velocity precisely, the first-order galactic differential rotation correction (DRC) was taken into account (Mihalas \& Binney, 1981), and -10.33 and -0.90 km s$^{-1}$ DRCs were applied to $u$ and $v$ space velocity components, respectively. The $w$ velocity is not affected in this first-order approximation. As for the local standard of rest correction, Mihalas \& Binney's (1981) values (9, 12, 7) km s$^{-1}$ were used and the total space velocity of LT CMa was obtained as $S=12.45$(3.76) km s$^{-1}$. To determine the population type of LT CMa, the galactic orbit of the system was examined. Using the N-body code of Dinescu, Girardi \& van Altena (1999), the system's apogalactic ($R_{max}$) and perigalactic ($R_{min}$) distances were obtained as 9.61 and 8.35 kpc, respectively. Also, the maximum possible vertical separation of the system from the galactic plane is $|z_{max}|=|z_{min}|=30$ pc. The following formulae were used to derive the planar and vertical ellipticities:

\begin{equation}
e_{p}=\frac{R_{max}-R_{min}}{R_{max} + R_{min}},
\end{equation}

\begin{equation}
e_{v}=\frac{|z_{max}|+|z_{min}|}{R_{max}+R_{min}}.
\end{equation}

The planar and vertical ellipticities were calculated as $e_{p}=0.07$ and $e_{v}=0$. These values show that LT CMa is orbiting the Galaxy in a circular orbit and the system belongs to the young thin-disc population.

\subsection{Internal Structure}

The internal structure constant, (\textit{$\bar{k}_{\rm 2}$}), is a measure of the density concentration of the components of eclipsing binary stars with eccentric orbits. It is an important parameter of stellar evolution models. If the orbital eccentricity, the orbital and apsidal motion periods, the masses and radii of the components are known, the mean internal structure constant of the component stars in the system can be derived.

The observational mean value of the internal structure constant, ($\textit{$\bar{k}_{\rm 2,obs}$}$), is given by the following formula (Kopal 1978):

\begin{equation}
\overline{k}_{2,obs}=\frac{1}{c_{21}+c_{22}}\frac{P}{U},
\end{equation}
where $P$ and $U$ are the orbital and apsidal period, and the constants $c_{\rm 2i}$ ($\textit{i}$ = 1, 2):

\begin{equation}
c_{2i}=r_{i}^{5}[(\frac{\omega_{r,i}}{\omega_{k}})^{2}(1+\frac{M_{3-i}}{M_{i}})f(e)+\frac{15M_{3-i}}{M_{i}}g(e)],
\end{equation}
where

\begin{equation}
f(e)=\frac{1}{(1-e^{2})^{2}},
\end{equation}

\begin{equation}
g(e)=\frac{(8+12e^{2}+e^{4})f(e)^{2.5}}{8}.
\end{equation}
Here \textit{r$_{\rm i}$}, \textit{$M_{\rm i}$} and $\omega_{\rm r,i}$ are the relative radii, masses and the angular velocities of the axial rotation of the components, respectively, and $\omega_{\rm k}$ is the mean angular velocity of orbital motion. The first term in Eq. 6 represents the contribution of rotational distortions, and the second term corresponds to the tidal effects.

It is well known that the observed $\textit{$\bar{k}_{\rm 2,obs}$}$ contains not only the Newtonian contribution to apsidal motion, which is due to the stars being distorted extended shapes, but also a small amount of relativistic apsidal motion. The theory of General Relativity estimates the relativistic contribution to the observed rate as the following Einstein formula
(Gim\'{e}nez, 1985):

\begin{equation}
\dot{\omega}_{rel} = 5.45 \times 10^{-4} \frac{1}{1-e^2}
(\frac{M_1+M_2}{P})^{2/3}
\end{equation}
where $M_i$ denotes the masses of the components in solar mass and $P$ is the period of the orbit in days.

The individual internal structure constants, $\textit{$\bar{k}_{\rm 2i}$}$, must be combined using the equation:

\begin{equation}
\bar{k}_{2,theo}=\frac{c_{21}k_{21,theo}+c_{22}k_{22,theo}}{c_{21}+c_{22}}.
\end{equation}
to find the weighted average coefficient which is directly comparable to the observed value.

The observed apsidal motion period of LT CMa $\textit{U}$ = 69 $\pm$ 5 yr, corresponding to a total rate of $\dot{\omega}$ = 0.0252 $\pm$ 0.0020 deg cycle$^{-1}$ was obtained in section 5. The relativistic contribution to the apsidal motion calculated from general relativity is 0.00162 deg cycle$^{-1}$ or amounting to about 6\% of the observed rate. After correcting for this effect, an average internal structure constant of LT CMa is derived, with assumption of the periastron-synchronization, to be $\textit{log$\bar{k}_{\rm 2,obs}$}$=$-$2.53.

The theoretical internal structure constants for the components ($\log k_{\rm 2i,theo}$) are taken from the new evolutionary models of Claret (2004) with the standard chemical composition of ($X$, $Z$) = (0.70, 0.02). Among the tabulated values, interpolation for the masses and $\log g$ of the component stars of LT CMa yielded -2.23 and -2.18 for the primary and the secondary component, respectively. The average theoretical value of internal structure constant was computed as $\bar{k}_{\rm 2,theo}$=-2.22. This value seems significantly different from the observed value, in the sense that the component stars appear to be more concentrated in mass than theoretically predicted from the evolutionary models. There are several possible reasons for disagreement in the case of LT CMa. Perhaps the most important is the fact that the range of observations used to derive the apsidal motion period is small compared to the apsidal motion period.
\\ \\
\textbf{Acknowledgements}\\
This work has been supported by the Scientific and Technological Research Council of Turkey (T\"{U}B\.{I}TAK) under the Project Number: TBAG 149T449. Observations in this study are granted by The T\"{U}B\.{I}TAK National Observatory under the Project Number: 09ARTT150-431-0. We thank to anonymous referee who helped to improve the early version of the manuscript with his/her very useful comments.

\end{document}